\documentclass[aps,prc,showpacs,nofootinbib,twocolumn]{revtex4}
\usepackage{epsfig}
\usepackage{graphicx,amsmath}
\newcommand\ba{\begin{eqnarray}}
\newcommand\ea{\end{eqnarray}}

\newcommand{\be}{\begin{equation}}
\newcommand{\ee}{\end{equation}}

\newcommand{\bas}{\begin{eqnarray*}}
\newcommand{\eas}{\end{eqnarray*}}
\begin{document}
\title{\bf \large Search for Low Mass Exotic mesonic structures. Part I: experimental results}

\author{
B. Tatischeff$^{1,2}$\thanks{e-mail : tati@ipno.in2p3.fr}\\
$^{1}$CNRS/IN2P3, Institut de Physique Nucl\'eaire, UMR 8608, Orsay, F-91405\\
$^{2}$Univ. Paris-Sud, Orsay, F-91405, France\vspace{3.mm}\\
E. Tomasi-Gustafsson$^{3}$\thanks{e-mail : etomasi@cea.fr}\vspace*{1.mm}\\
$^{3}$DAPNIA/SPhN, CEA/Saclay\\ 91191 Gif-sur-Yvette Cedex, France}

%\date{ }

\vspace*{1cm}
\begin{abstract}
Recently, several papers discussed on the existence of a low mass new structure at a mass close to M=214.3~MeV. It was suggested that the $\Sigma^{+}$ disintegration: 
$\Sigma^{+}\to$pP$^{0}$, P$^{0}\to\mu^{-}\mu^{+}$ proceeds through an intermediate particle
P$^{0}$ having such mass.
The present work intends to look at other new or available data, in order to observe the eventual existence of small narrow peaks or shoulders in very low mesonic masses. Indeed narrow structures were already extracted from various data in dibaryons, baryons and mesons (at larger masses that those studied here).
\end{abstract}
\maketitle
\section{Introduction} 
The $\Sigma^{+}$ disintegration: $\Sigma^{+}\to$pP$^{0}$, P$^{0}\to\mu^{-}\mu^{+}$  was studied at Fermilab by H. Park {\it et al.}  \cite{park}. The data were taken by the HyperCP (E871) Collaboration. The authors observed a narrow range of dimuon masses, and supposed that the decay may proceed via a neutral intermediate state P$_{0}$, with a mass M=214.3~MeV $\pm$ 0.5~MeV. 

Several theoretical works were done assuming the existence of this new particle. He, Tandean, and Valencia performed a standard-model interpretation of the data \cite{he1}. Later on the same authors demonstrate that the new particle could be a pseudoscalar or axial-vector, but not scalar nor vector \cite{xiao}. They also suggested that the particle could be a very light pseudoscaler Higgs \cite{he2}. Deshpande 
{\it et al.}  \cite{desh} assume a fundamental spin zero boson, which couple to quarks with flavor changing transition s$\to$d$\mu^{+}\mu^{-}$. They estimate the scalar and pseudoscalar coupling constants and evaluate several branching ratios. Geng and Hsiao \cite{geng} found that the P$^{0}$ cannot be scalar but pseudoscalar, and determine that the decay width should be as small as $\approx$10$^{-7}$MeV. Gorbunav and Rubakov \cite{gorb} discuss possible sgoldstino interpretation of this possible particle.

The experimental observation was based on three events. We anticipate that this low counting is due to their observation in a weak disintegration channel.
In order to eventually strengthen this result by a direct observation, we look at already existing data and try to observe a possible signature of (a) small peak(s), or (a) small shoulder(s), (at a mass not far from the mass of P$^{0}$).
Such mass(es) can be observed, either in the invariant masses of two muons, M$_{\mu\mu}$, or in missing masses of different reactions, studied  with incident leptons as  well as with incident hadrons. However the signal, if any, is expected to be small. The spectra are therefore presented in the semi-log scale. The signals will be superposed to a relatively large tail of one pion missing mass. Therefore the signal, if any, can only be observed in precise data, with large statistics, good resolution and small binning. Moreover, the mass range studied must be small. Such data are scarce and concern reactions studied at rather low incident energies, with good resolution. When we found a hint for a small effect, we read out  and reanalyzed the data. Several such structures were selected and presented below. 
%%%%%%%%%%%%%%%%%%%%%%%%%%%%%%%%%%%%%%%%%%%%%%%%%%%%%%%%%%
\section{Selected data showing small structures in the mass range above the pion mass}
%180$\le$M$\le$240~MeV}
%%%%%%%%%%%%%%%%%%%%%%%%%%%%%%%%%%%%%%%%%%%%%%%%%%%%%%%%%%%
\subsection{The missing mass of the pp$\to$ppX reaction}
%%%%%%%%%%%%%%%%%%%%%%%%%%%%%%%%%%%%%%%%%%%%%%%%%%%%%%%%%%%%%%%%%%%

The pp$\to$ppX (X -'meson type') reaction was studied at SATURNE (SPES3 beam line), at T$_{p}$=1520, 1805, and 2100~MeV \cite{bt2}. The missing mass displays a broad structure, in the mass range
280$\le$M$\le$580~MeV, unstable for different kinematical conditions and slightly oscillating \cite{jy}, previously called the ABC effect; it was analysed as being due to a superposition of four narrow mesonic states: M=310~MeV, 350~MeV, 430~MeV, and 495~MeV \cite{jy}. Above the $\eta$ mass, narrow mesonic structures were extracted at the following masses: M=550, 588, 608, 647, 681, 700, 715, and 750~MeV \cite{bt1}.

Since the widths of the missing mass peaks increase for increasing spectrometer angles, we keep only the three lowest angle spectra at T$_{p}$=1520~MeV, add them, and show the resulting spectra in 
Fig.~1(a). The high counting rate allows to extract a clear peak at M$_{X}$=216.5~MeV. 

The reaction pp$\to$pp$\pi^{0}\pi^{0}$ was studied close to threshold at CELSIUS (Uppsala) \cite{bilger}. The missing mass data, after integration over two channels, are shown in Fig.~1(b). The two $\pi^{0}$ phase space starts at M$_{X}\approx$240~MeV.  The events in the range 170~$\le$M$_{X}\le$ 240~MeV, are mostly physical as the  background contribution is estimated to less than 10 events/channel.
These data are fitted with a $\pi^{0}$ peak and two small structures, having the same shape as the $\pi^{0}$ peak, are extracted at M$_{X}$=182 and 220~MeV.

The pp$\to$ppX reaction was also studied at J$\ddot{u}$lich COSY-TOF \cite{roder}.  Both protons in the final state were detected in order to study the $\eta$ production. The data were read and shown in Fig.~1(c). Since they are given in the original work as a function of the missing mass squared with constant binning ($\Delta$M$_{X}$=0.002~GeV$^{2}$), they are plotted versus M$_{X}$ as given, up to 210~MeV (empty circles), and for larger missing mass they are integrated over two channels (full circles). The peak corresponding to $\pi^{0}$ missing mass is fitted by a gaussian, at M$_{X}$=135~MeV ($\sigma$=67~MeV), and the data at larger missing mass are fitted with a polynomial. Two structures can be extracted, the first one at M=197~MeV, not valid statistically, and the second at M=224~MeV. They are very narrow, therefore, if fitted by only one structure, which includes both narrow structures, they result in a broad gaussian centered at M=214~MeV (dashed curve in Fig~1(c))
\begin{figure}[!ht]
%219.eps dans home (219.kumac dans depou1)
\begin{center}                                                          
\scalebox{.45}[.5]{
\includegraphics[bb=33 23 520 520,clip=]{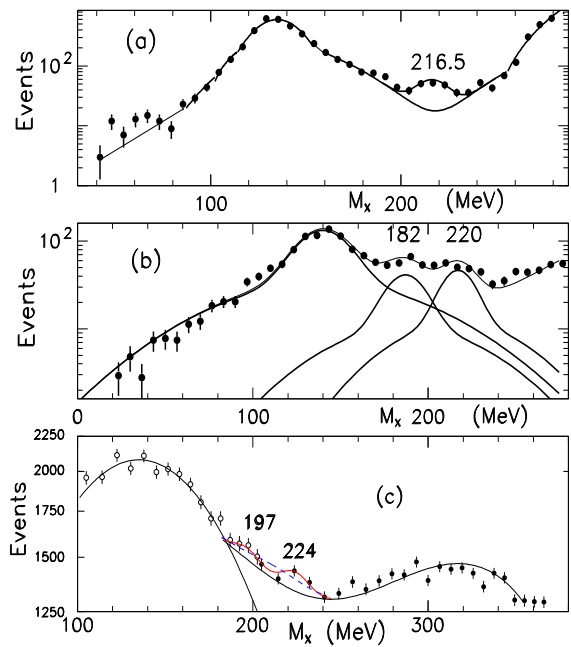}}
\caption{Insert (a): missing mass of the pp$\to$ppX reaction measured at SATURNE (SPES3 beam line) at T$_{p}$=1520~MeV. Three spectra measured at $\theta_{pp}$=0$^{0}$, 2$^{0}$, and 5$^{0}$, are added. Insert (b): same reaction measured at CELSIUS  \protect\cite{bilger}. Insert (c): Missing mass of the pp$\to$ppX reaction studied at J$\ddot{u}$lich COSY-TOF \protect\cite{roder}.}
\label{Fig1}
\end{center}
\end{figure}
\begin{figure}[!ht]
%weis.kumac , weiss.eps
\begin{center}                                                          
\scalebox{.42}[.5]{
%\scalebox{.25}[.25]{
\includegraphics[bb=44 24 530 530,clip=]{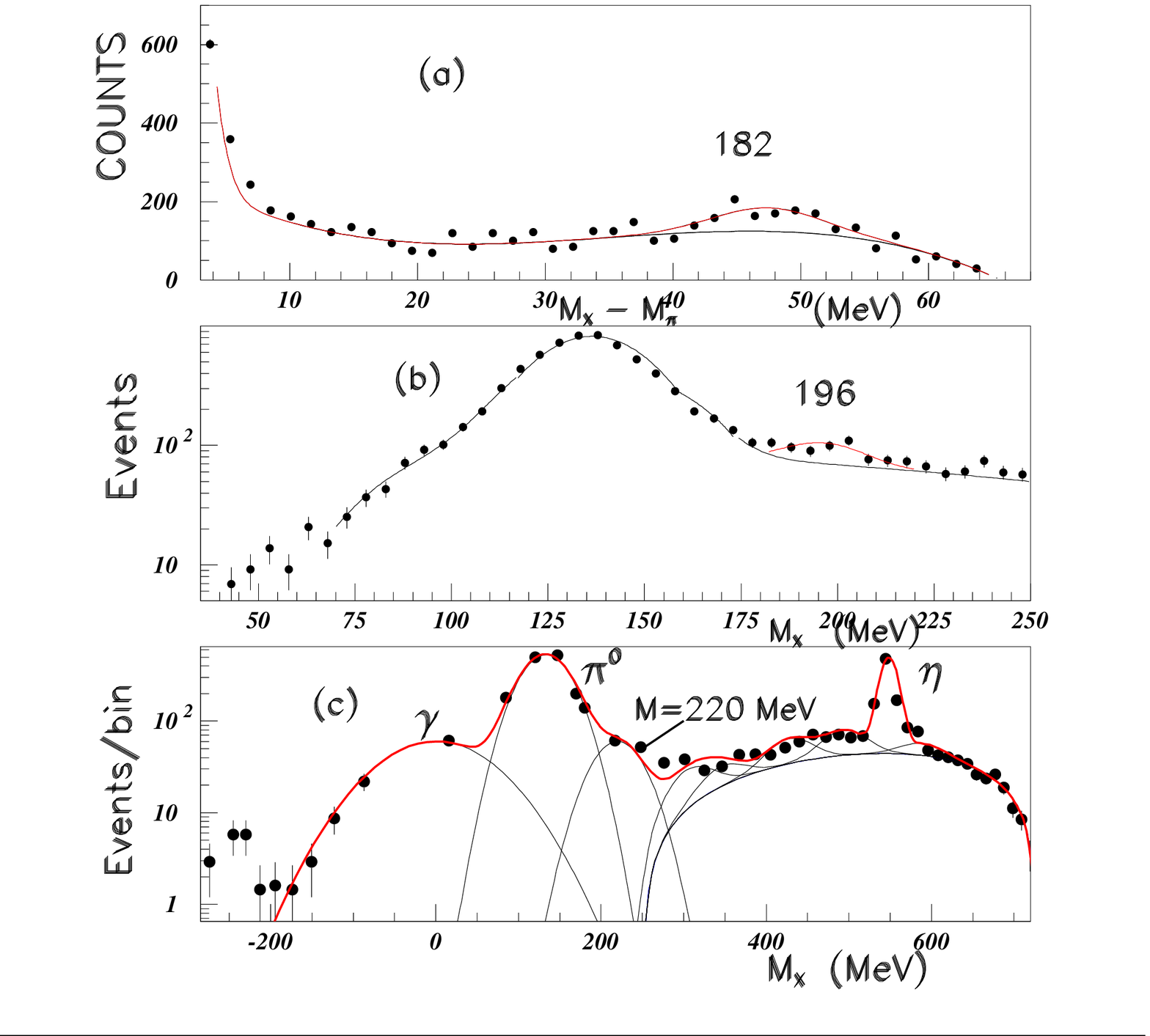}}
\caption{Insert (a): $\pi^{0}$ electroproduction at threshold, measured at MAMI \protect\cite{weis}. The missing mass spectra is integrated over 4 channels. 
 Insert (b): missing mass of the p($\vec{e},e'\vec{p})\pi^{0}$ reaction studied at JLAB Hall A at $\theta_{cm}$=90$^{0}$ \protect\cite{sirca}. 
Insert (c): missing mass of the p(e,e'p)X reaction measured at JLAB Hall C \protect\cite{frolov}.}
\label{Fig2}
\end{center}
\end{figure}
\subsection{The missing mass of the ep$\to$e'pX reaction}
The $\pi^{0}$ electroproduction at threshold  for Q$^{2}$=0.05~GeV$^{2}$ was measured at MAMI \cite{weis}. The missing mass spectrum, up to M$_{X}$=200~MeV is given in Fig.~3(b) of \cite{weis}, after background subtraction. The data are read, integrated over 4 channels and reported in Fig.~2(a). A peak at M$_{X}$=182~MeV is clearly observed. Indeed, the resolution in these data is as good as FWHM=2.2~MeV, as given in the $\pi^{0}$ peak (removed here to enhance the mass range discussed). The increase of the number of events between 48$\le$M$_{X} $~- M$_{\pi}\le$56~MeV is physical. The contribution from two pion production, cannot be large at a so low mass value as M=180~MeV. 

The Roper resonance was studied at JLAB Hall A using  the p($\vec{e},e'\vec{p})\pi^{0}$ reaction \cite{sirca}. Two missing mass spectra were given at $\theta_{cm}$=90$^{0}$ and $\theta_{cm}$=-90$^{0}$. No shoulder is observed in this last spectra. The values of the spectra at $\theta_{cm}$=90$^{0}$ are read and shifted in order to put the $\pi^{0}$ peak at his right mass, namely at M$_{X}$=135~MeV. Fig.~2(b) shows this spectrum fitted with a gaussian and two polynomials. A  small enhancement is observed at M$_{X}$=196~MeV.

The $\pi^{0}$ electroproduction on the proton was studied in Hall C at JLAB \cite{frolov}, in the region of the $\Delta$(1232) resonance via the  p(e,e'p)$\pi^{0}$ reaction. The authors give in Fig.~1 of Ref. \cite{frolov} an example of missing mass distribution for the reaction p(e,e'p)X. These data are read and reported in Fig.~2(c). The widths of all $\pi^{0}$ and $\eta$ peaks are related to their masses (proportional to 1/M). These widths define the width of the small peak extracted at M=220~MeV. Several other peaks are introduced, following the results of the pp$\to$ppX reaction studied at SPES3 (SATURNE, Saclay) \cite{jy}. After introduction of an arbitrary two-pion phase space, a contribution of the p(e,e'p)$\gamma$ reaction is observed around M$_{X}$=0.
\begin{figure}[!h]
%prl7f3.kumac
\begin{center}                                                          
\scalebox{.47}[.5]{
\includegraphics[bb=20 20 530 530,clip=]{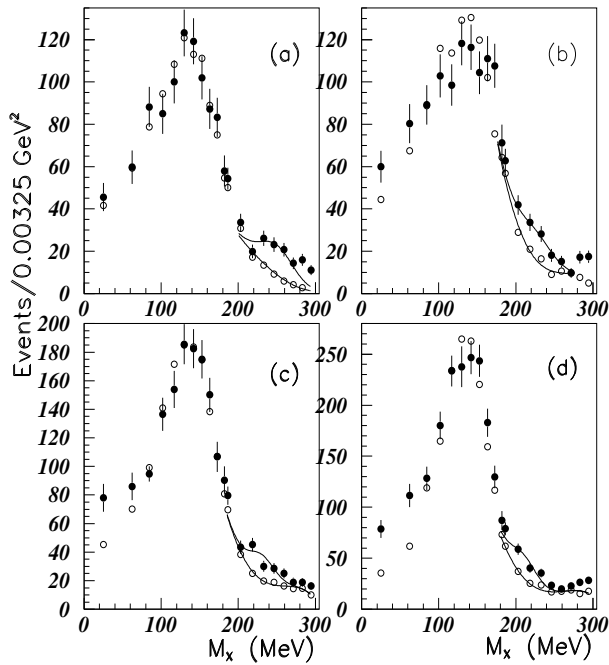}}
\caption{The missing mass of the  p(e,e'p)X reaction \protect\cite{frolovt} studied at JLAB Hall C at Q$^{2}$=4.0~GeV$^{2}$. The empty circles correspond to Monte-Carlo simulations; the full circles correspond to data. Inserts (a), (b), (c), and (d) correspond respectively to  p$^{0}_{p}$=2~GeV and $\theta^{0}_{p}$=23$^{0}$, 
 p$^{0}_{p}$=2~GeV and $\theta^{0}_{p}$=20$^{0}$, p$^{0}_{p}$=2.2~GeV and $\theta^{0}_{p}$=17$^{0}$, and 
 p$^{0}_{p}$=2.45~GeV and $\theta^{0}_{p}$=17$^{0}$.}
\label{Fig3}
\end{center}
\end{figure}
More detailed data from the same experiment \cite{frolovt}, are reported in  several spectra where structures can be extracted in the same missing mass range. The measurements were performed for two values of the four momentum transfer squared between the initial and the final electron, namely at Q$^{2}$=2.8~GeV$^{2}$ and Q$^{2}$=4~GeV$^{2}$.  The measurements were performed for a few values of P$^{0}_{p}$ and a few values of $\theta^{0}_{p}$. Four spectra are shown in Fig.~3 for  Q$^{2}$=4~GeV$^{2}$. 
Fig.~4 shows another selection of spectra corresponding to Q$^{2}$=2.8~GeV$^{2}$. In both figures empty circles correspond to Monte-Carlo simulations \cite{frolovt} and full circles correspond to data. In the missing mass range studied here an excess of counts can be seen between data and the simulation which were fitted by a polynomial. The quantitative informations are given in Table~I. The discrepancy between data and simulation for M$_{X}\le$60~MeV  has been attributed to the Bethe-Heitler process (ep$\to$e'p'$\gamma$ reaction).

\begin{figure}[!ht]
%prl7f4.kumac
\begin{center}                                                          
\scalebox{.47}[.5]{
\includegraphics[bb=20 20 530 530,clip=]{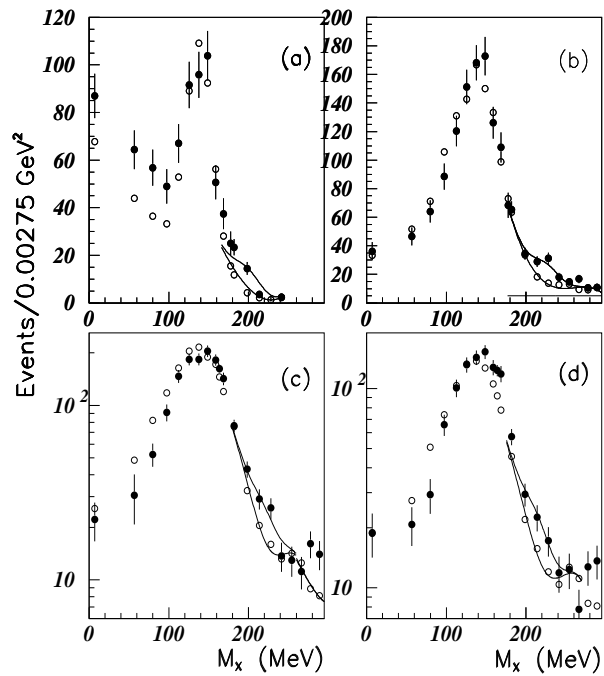}}
\caption{The missing mass of the  p(e,e'p)X reaction \protect\cite{frolovt} studied at JLAB Hall C at Q$^{2}$=2.8~GeV$^{2}$. The empty circles correspond to Monte-Carlo simulations; the full circles correspond to data. Inserts (a), (b), (c), and (d) correspond respectively to  p$^{0}_{p}$=1.9~GeV and $\theta^{0}_{p}$=33$^{0}$, 
 p$^{0}_{p}$=1.55~GeV and $\theta^{0}_{p}$=23$^{0}$, p$^{0}_{p}$=1.7~GeV and $\theta^{0}_{p}$=19$^{0}$, and 
 p$^{0}_{p}$=1.7~GeV and $\theta^{0}_{p}$=23$^{0}$.}
\label{Fig4}
\end{center}
\end{figure}
\vspace*{-3.mm}
%%%%%%%%%%%%%%%%%%%%%%%%%%%%%
\subsection {Discussion}
%%%%%%%%%%%%%%%%%%%%%%%%%%%%%
We have looked at some existing data, in order to find evidence for the existence of a new boson.
All spectra shown here, display a structure, but at slightly different masses. However, there is an indication of a possible regrouping around several mass values. The statistics is too low for giving an evidence if the results privilege one unstable mass or a few better defined masses.  We increase therefore the number of spectra studied, as those shown in Figs~3 and 4. These spectra are not shown here. The corresponding quantitative informations are summarized in Table~I. They favor a regrouping into several values; the same conclusion is favored by the existence of more than one peak in the same spectrum, as in Fig.~1(b). In summary these narrow structures masses (see Fig.~5), are tentatively observed at:\\
\hspace{3.mm}M=181$\pm$2~MeV (5 events),\\
\hspace{3.mm}M=198$\pm$2~MeV (5 events),\\
\hspace{3.mm}M=215$\pm$5~MeV (12 events),\\
\hspace{3.mm}M=227.5$\pm$2.5~MeV (5 events),\\
\hspace{3.mm}M=235$\pm$1~MeV (3 events).\\
\begin{figure}[!h]
%prl7f5.kumac
\begin{center}                                                          
\scalebox{.47}[.5]{
\includegraphics[bb=47 284 520 520,clip=]{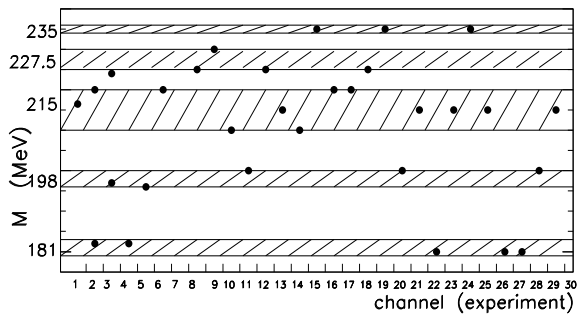}}
\caption{Masses of the weakly excited structures extracted from several experiments (see text and table~I).}
\label{Fig5}
\end{center}
\end{figure}
We notice that the range exhibiting the largest number of experimental mass structures, namely around M=215~MeV, agrees with the value extracted at Fermilab: M=214.3~MeV \cite{park}.  There is also an additionnal but qualitative evidence in favour of a structure at M$\approx$214~MeV. The pd$\to$pd$\eta$ reaction was studied at CELSIUS  \cite{bilg}.
Fig.~4 of Ref. \cite{bilg} (lower frame) shows a scatterplot of M$_{\gamma\gamma}$ versus M$_{pd}$,
where a careful observation indicates an excess of counts around M$_{\gamma\gamma}\approx$214~MeV. Such resonances could decay into two electrons, but the probability of the decay into two leptons is proportional to their mass, so it is strongly disfavored with respect to muons.
\begin{table}[t]
\caption{Masses (in MeV) and number of standard deviations (S.D.) of the narrow peaks extracted around M$\approx$215~MeV, from the p(e,e'p)X reaction studied at JLAB Hall C \protect\cite{frolovt} for Q$^{2}$=4~GeV$^{2}$ (Fig.~3) and Q$^{2}$=2.8~GeV$^{2}$ (Fig.~4). The width of the peak is given by $\sigma$ (in MeV).  p$^{0}_{p}$ is in GeV, and $\theta$ is in degrees.}
\label{Table I}
\vspace{5.mm}
\begin{tabular}{c c c c c c c c c c}
\hline
Fig.&insert&set in \protect\cite{frolovt}&Q$^{2}$&p$^{0}_{p}$&$\theta^{0}_{p}$&M&$\sigma$&S.D.\\
\hline
Fig.~3&(a)&10&4.0&2&23&250&22&9.5\\
          &(b)&11&4.0&2&20&225&22&4.9\\
          &(c)&21&4.0&2.2&17&230&17&5.2\\
          &(d)&5&4.0&2.45&17&210&17&6.3\\   
\hline
Fig.~4&(a)&14&2.8&1.9&33&200&17&1.9\\
            &(b)&19&2.8&1.55&23&225&17&4.0\\  
            &(c)&34&2.8&1.7&19&215&17&3.0\\          
            &(d)&36&2.8&1.7&23&210&17&2.9\\
\hline
\hline
&&12&4.0&2&17&235&22&5.1\\
&&14&4.0&1.8&17&220&22&2.6\\
&&15&4.0&1.8&20&220&17&4.9\\
&&16&4.0&1.8&23&225&17&1.5\\
&&19&4.0&2.2&23&235&17&5.3\\
&&20&4.0&2.2&20&200&17&5.9\\
&&4&4.0&2.45&14&215&17&3.5\\
&&6&4.0&2.45&20&180&17&3.7\\   
\hline
&&9&2.8&1.9&23&215&17&3.8\\
&&10&2.8&1.9&25&210&17&3.75\\
&&13&2.8&1.9&31&235&17&1.9\\
&&18&2.8&1.55&25&215&17&3.5\\
&&28&2.8&2.15&23&180&17&11\\
&&29&2.8&2.15&21&180&17&8.6\\
&&30&2.8&2.15&19&200&17&3.4\\
&&33&2.8&1.7&17&215&17&2.8\\            
\hline
\hline
\end{tabular}
\end{table}

%%%%%%%%%%%%%%%%%%%%%%%%%%%%%%%%%%%%%%%%%%%%%%%%%%%%%%%%%%
\section{Selected data showing small structures in the mass range below the pion mass}
%%%%%%%%%%%%%%%%%%%%%%%%%%%%%%%%%%%%%%%%%%%%%%%%%%%%%%%%%%%%%%%%%%%%%%%%%
\subsection{The missing mass of the pp$\to$ppX reaction}
%%%%%%%%%%%%%%%%%%%%%%%%%%%%%%%%%%%%%%%%%%%%%%%%%%%%%%%%%%%%%%%%%%%%%%%%%
The natural question, following the previous result, is to look, in already published data, at (a) possible structure(s) below the mass of the pion (M=135~MeV). All the data reanalyzed below, are read and their mass recalibrated, when necessary, to adjust the pion peak at M=135~MeV. The pion missing mass peak is described by a gaussian and the structure(s) at lower mass(es) is (are) described by a gaussian with the same width as the one given for $\pi^{0}$.
The observed structures are small, therefore a semi-log scale is used. Also the mass(es) extracted is (are) not always stable since the corresponding statistics is not reach. A small background is arbitrarily drawn. If it will be modified, the results will not change much, since we are in the xesemi-log scale.

 Fig.~6 shows the missing mass spectra studied at SATURNE (SPES3 beam line) in the useful range, at T$_{p}$=1520~MeV, and at four different spectrometer angles. A small peak is easily extracted at forward angles. When the spectrometer angle increases, the excitation of this exotic structure increases also relatively to the $\pi^{0}$ excitation, but the resolution gets spoiled and the peak although still extracted,  is no more clearly separated from the $\pi^{0}$ peak. Fig.~7 shows another selection of four missing mass spectra.
\begin{figure}[h]
%75.kumac (depou1)
\begin{center}                                                          
\scalebox{.48}[.48]{
\includegraphics[bb=20 20 530 530,clip=]{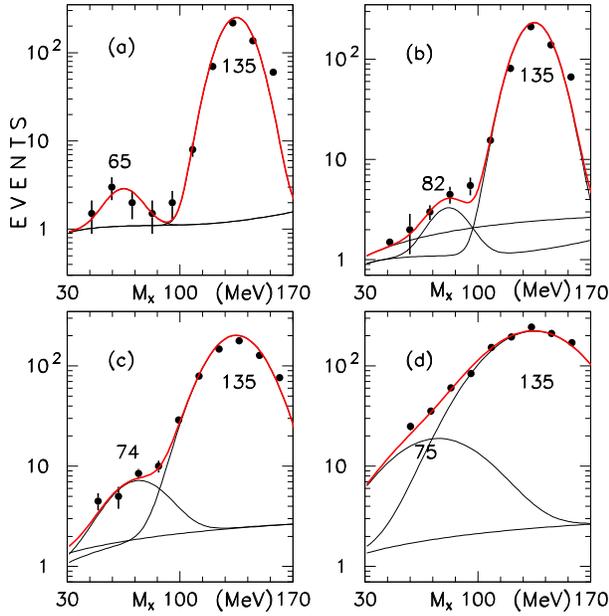}}
\caption{Missing mass spectra for pp$\to$ppX measured at SATURNE (SPES3 beam line) at T$_{p}$=1520~MeV. Inserts (a), (b), (c), and (d) correspond respectively to the following spectrometer angles:  $\theta_{spec.}$=0$^{0}$,  2$^{0}$, 5$^{0}$, and 9$^{0}$.}
\label{fig6}
\end{center}
\end{figure}
\begin{table}[t]
\caption{Quantitative information on the small structure extracted from the missing mass spectra studied with pp$\to$ppX reaction at SATURNE (SPES3 beam line) \protect\cite{bt2}. The incident proton energies T$_{p}$ and the mass M$\approx$65~ are in MeV. R is the ratio of the M$\approx$65~MeV
structure excitation over the $\pi^{0}$ excitation.}
\label{Table II}
\vspace{5.mm}
\begin{tabular}{c c c c c c}
\hline
Fig.&T$_{p}$&$\theta_{pp}$&$\sigma$&M$\approx$65&R\\
\hline
Fig.~6(a)&1520&0&10.3&65&7.2 10$^{-3}$\\
Fig.~6(b)&1520&2&11.5&82&9.6 10$^{-3}$\\
Fig.~6(c)&1520&5&17&74&27 10$^{-3}$\\
Fig.~6(d)&1520&9&28&75&77 10$^{-3}$\\
\hline
Fig.~7(a)&1520&13&38&60&9.4 10$^{-2}$\\
Fig.~7(b)&1520&17&38&60&18.7 10$^{-2}$\\
Fig.~7(c)&1805&9&27&85&13.8 10$^{-2}$\\
Fig.~7(d)&1805&13&40&60&10.0 10$^{-2}$\\
\hline
\hline
\end{tabular}
\end{table}
Here again the experimental results are well fitted with introduction of a second peak at a mass close to M$\approx$65~MeV. Table~II gives the quantitative informations. $\sigma$ describes the width of the peaks  and R is the ratio of the exotic structure excitation relative to the $\pi^{0}$ excitation.
\begin{figure}[t]
%76.kumac (depou1)
\begin{center}                                                          
\scalebox{.48}[.48]{
\includegraphics[bb=20 20 530 530,clip=]{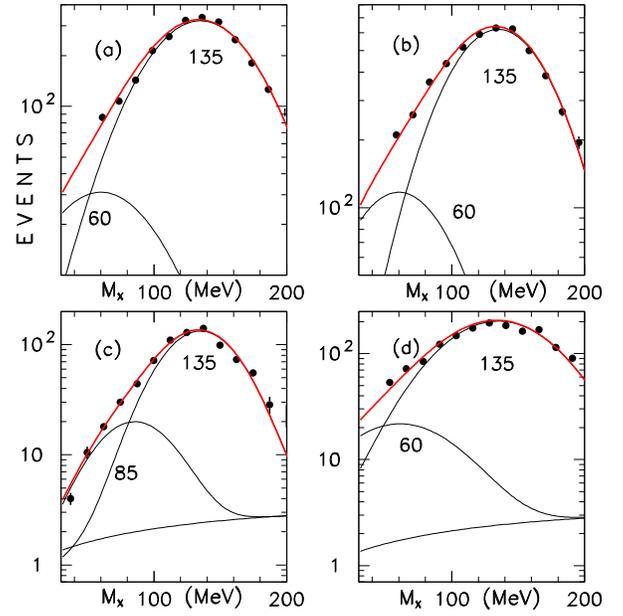}}
\caption{Missing mass spectra for pp$\to$ppX measured at SATURNE (SPES3 beam line). Inserts (a), (b), (c), and (d) correspond respectively to the following kinematical conditions: 
T$_{p}$=1520~MeV, $\theta_{spec.}$=13$^{0}$;
T$_{p}$=1520~MeV, $\theta_{spec.}$=17$^{0}$;
T$_{p}$=1805~MeV, $\theta_{spec.}$=9$^{0}$; and
T$_{p}$=1805~MeV, $\theta_{spec.}$=13$^{0}$.}
\label{fig7}
\end{center}
\end{figure}

Fig.~8 shows a selection of missing mass peaks from CELSIUS in inserts (a), (b), and (c). Insert (a) shows the data from the pp$\to$pp$\pi^{+}\pi^{-}$ reaction studied at CELSIUS  \cite{patz} at T$_{p}$=775~MeV. The $\sigma$ of the peaks equals 20~MeV, and R=17~10$^{-2}$. Insert (b) shows the data from the pp$\to$pp$\gamma\gamma$ reaction studied at CELSIUS  \cite{kull} at T$_{p}$=1360~MeV. Here 
$\sigma$=16~MeV and R=6.6~10$^{-2}$. Insert (c) shows the data of the 
 pp$\to$ppX reaction studied at CELSIUS  \cite{bilger} at T$_{p}$=650~MeV. Here $\sigma$=17~MeV and 
R=8.3~10$^{-2}$ for the ratio of the "60"/"135" peaks and R=20~10$^{-2}$ for the ratio of the "100"/"135" peaks. In all these spectra a peak at M$\approx$65~MeV is observed, and also another one is extracted at M=100~MeV. Table~II gives the quantitative informations.
\begin{figure}[h]
%patz1.kumac
\begin{center}                                                          
\scalebox{.48}[.48]{
\includegraphics[bb=20 20 530 530,clip=]{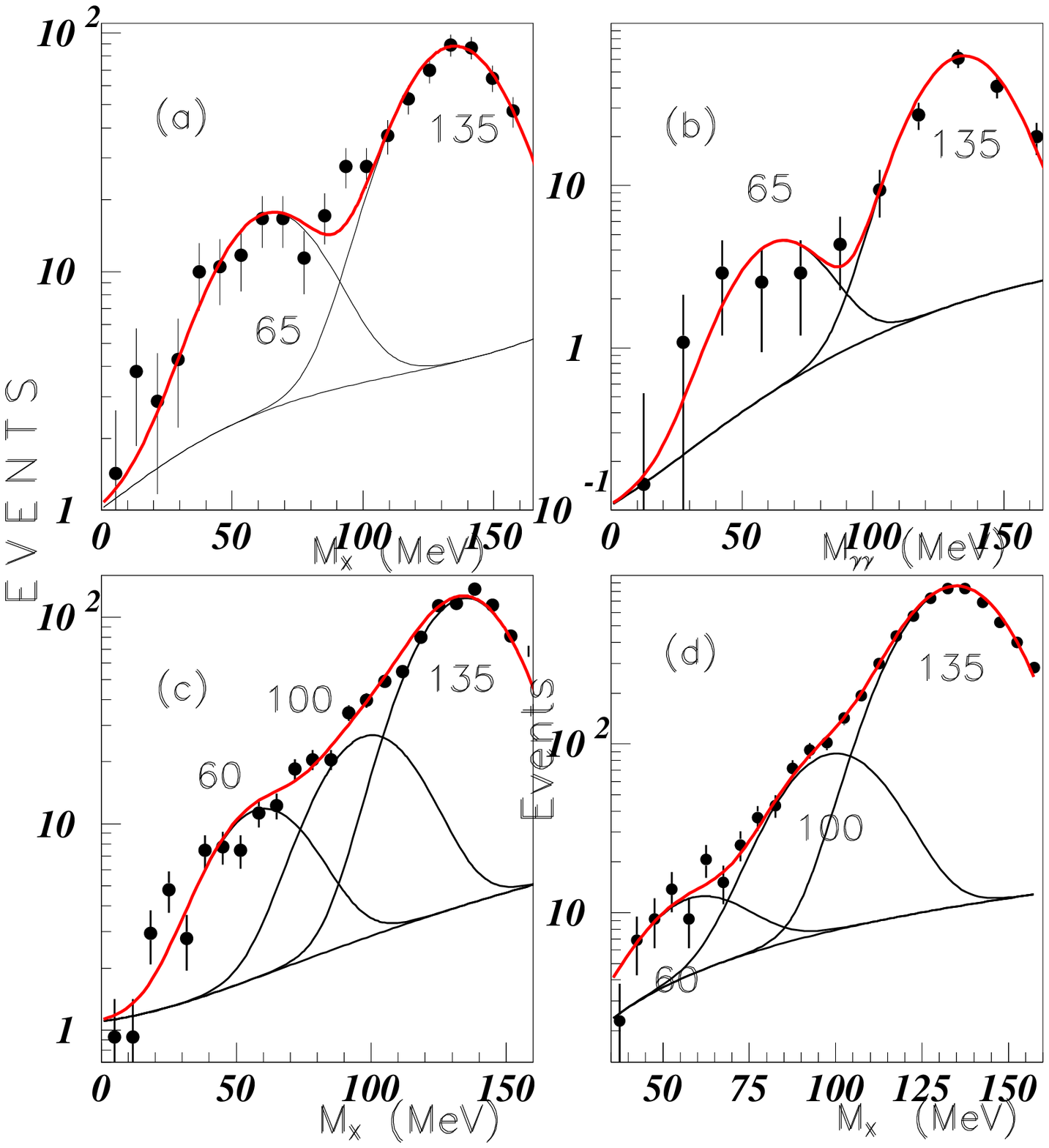}}
\caption{Missing mass spectra for several different reactions measured at CELSIUS . (a) pp$\to$pp$\pi^{+}\pi{-}$ \protect\cite{patz}, (b) pp$\to$pp$\gamma\gamma$
\protect\cite{kull}, (c) pp$\to$ppX \protect\cite{bilger}. Insert (d) shows the missing mass of the p($\vec{e},e'\vec{p})\pi^{0}$ reaction \cite{sirca} studied at JLAB Hall A at $\theta_{cm}$=90$^{0}$.} 
\label{fig8}
\end{center}
\end{figure}
\begin{figure}[b]
%smyrski.kumac (home)
\begin{center}                                                          
\scalebox{.48}[.48]{
\includegraphics[bb=20 20 530 530,clip=]{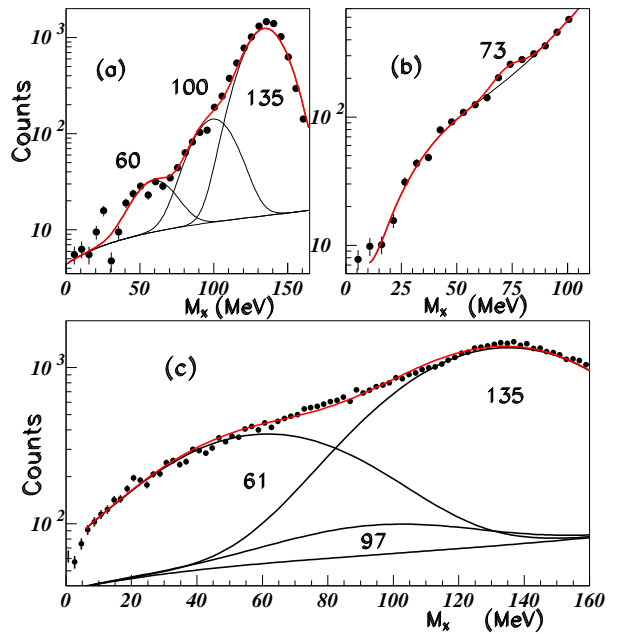}}
\caption{Missing mass spectra for several different reactions measured at COSY. (a) pd$\to^{3}$He$\pi^{0}$ \protect\cite{beti}, (b) pd$\to$T$\pi^{+}$ \protect\cite{mach},
(c) dp$\to^{3}$He$\eta$ \protect\cite{smyrski}. } 
\label{fig9}
\end{center}
\end{figure}
\begin{figure}[!h]
%jlpd.kumac
\begin{center}                                                          
\scalebox{.48}[.48]{
\includegraphics[bb=20 20 530 530,clip=]{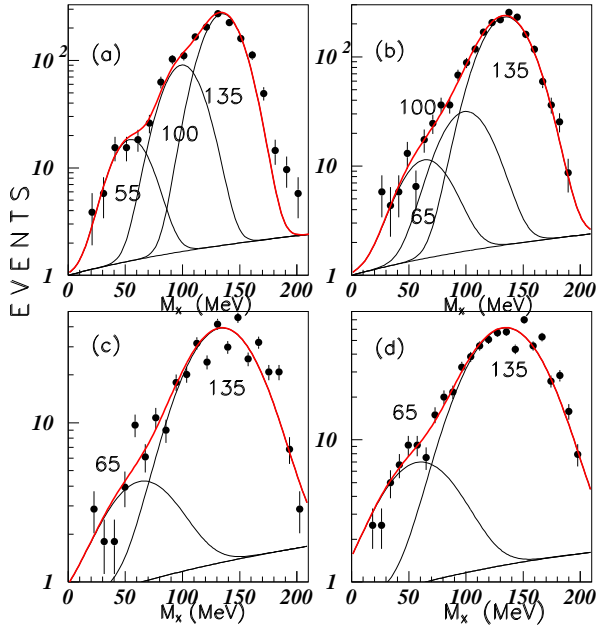}}
\caption{Missing mass spectra for $\gamma$p$\to$pX measured at JLAB Hall B \protect\cite{vineyard} CLAS Collaboration. Inserts (a), )b), (c), and (d) correspond respectively to the following kinematical conditions: 0.8$\le$E$\le$0.9~GeV and -0.75$\le$cos(T)$\le$-0.50, 0.8$\le$E$\le$0.9~GeV and -0.25$\le$cos(T)$\le$-0.00,
1.2$\le$E$\le$1.3~GeV and -0.50$\le$cos(T)$\le$-0.25, and
1.2$\le$E$\le$1.3~GeV and 0.50$\le$cos(T)$\le$0.75. E is the beam energy and T is the proton center-of-mass angle.}
\label{fig10}
\end{center}
\end{figure}
\begin{table}[t]
\caption{Quantitative informations concerning figs.~8, 9, and 10.}
\label{Table III}
\vspace{5.mm}
\begin{tabular}{c c c c c c}
\hline
Fig.&reaction&ref&lab.&M (R)&M (R)\\
\hline
Fig.~8(a)&pp$\to$pp$\pi^{+}\pi^{-}$&\protect\cite{patz}&CELSIUS &&65 (0.17)\\
Fig.~8(b)&pp$\to$pp$\gamma\gamma$&\protect\cite{kull}&CELSIUS &&65 (0.07)\\
Fig.~8(c)&pp$\to$pp$\pi^{0}\pi^{0}$&\protect\cite{bilger}&CELSIUS &100 (0.2)&60 (0.08)\\
Fig.~8(d)&p($\vec{e},e'\vec{p})\pi^{0}$&\protect\cite{sirca}&JLAB A&100 (0.09)&60 (0.01)\\
\hline
Fig.~9(a)&pd$\to^3$He$\pi^{0}$&\protect\cite{beti}&COSY&100 (0.1)&60 (0.02)\\
Fig.~9(b)&pd$\to$T$\pi^{+}$&\protect\cite{mach}&COSY&&73 \\
Fig.~9(c)&dp$\to^{3}$He$\eta$&\protect\cite{smyrski}&COSY&97 (0.03)&61(0.25)\\
\hline
Fig.~10(a)&$\gamma$p$\to$pX&\protect\cite{vineyard}&JLAB B&100 (0.33)&55 (0.06)\\
Fig.~10(b)&$\gamma$p$\to$pX&\protect\cite{vineyard}&JLAB B&100 (0.13)&65 (0.04)\\
Fig.~10(c)&$\gamma$p$\to$pX&\protect\cite{vineyard}&JLAB B& &65 (0.09)\\
Fig.~10(d)&$\gamma$p$\to$pX&\protect\cite{vineyard}&JLAB B& &65 (0.10)\\
\hline
\hline
\end{tabular}
\end{table}
Several spectra from COSY-Julich are reported in Fig.~9. They are all integreated by two channels in order to increase the precision. The effect in the spoiling of the resolution is observed, going from insert (a) to insert (c). Table~III gives the quantitative informations.
 Insert (a) shows the data from the pd$\to^{3}$He$\pi^{0}$ reaction measured by the GEM detector at COSY \cite{beti} at T$_{p}$=328~MeV. Here $\sigma$=13~MeV and R=1.7~10$^{-2}$ for the ratio of the "60"/"135" peaks and R=10.9~10$^{-2}$ for the ratio of the "100"/"135" peaks.
A large statistics missing mass spectra was obtained with the pd$\to^{3}$T$\pi^{+}$ reaction studied at COSY \cite{mach} at T$_{p}$=328~MeV also. The authors said that "small background was subtracted for each angular bin". The data are read, integrated by two channels, and shown in Fig.~9(b). A small peak at M=73.6~MeV is extracted. Insert (c) shows the missing mass spectra of the dp$\to^3$He$\eta$ reaction at T$_{d}$=1780~MeV \cite{smyrski}.
Here $\sigma$=29~MeV and R=25.2~10$^{-2}$ for the ratio of the "61"/"135" peaks and R=2.8~10$^{-2}$ for the ratio of the "97"/"135" peaks.

The missing mass of the p($\vec{e},e'\vec{p})\pi^{0}$ reaction \cite{sirca} studied at JLAB Hall A at $\theta_{cm}$=90$^{0}$ is read and reported in Fig.~8(d). Two structures, at M=100~MeV and 60~MeV are extracted.
\subsection{The missing mass of the $\gamma$p$\to$pX reaction}
The missing mass of the $\gamma$p$\to$pX reaction was studied at JLAB
in Hall B, in an experiment devoted to study the inclusive $\eta$ photoproduction in nuclei \cite{vineyard} by the CLAS Collaboration. The data at low missing mass range are read and reported in Fig.~10. We observe the good fit obtained with introduction of a structure at M=100~MeV in inserts (a) and (b) and a structure at M=65~MeV (M=55~MeV in insert (a)) in all inserts.

The mean values of the two low mass structures extracted from the various spectra shown, are M=62~MeV and M=100~MeV.
\section{Conclusion}
 Fig.~11 shows the various exotic masses shown in previous figs.. These masses are M=62~MeV, 80~MeV, 100~MeV, 181~MeV, 198~MeV, 215~MeV, 227.5~MeV, and 235~MeV, although the last one may be uncertain, since determined by only three data, and being located at the limit of the spectra. A few points, located around M=75~MeV, may be thought as being not resolved structures. Indeed when they are extracted none of the structures at M=100 or M=62~MeV is observed. However the symmetry of the masses reported in Fig.~11, may be considered as an indication of their genuine existence. 

We have selected some spectra showing these structures. In many other spectra, such extraction is not possible, either since their experimental resolution is worse, either since the dynamics of the experiment (reaction, incident energy ...) is less favourable. In figs.~6 and 7, and table~II, six spectra obtained at the same incident energy and same reaction, show that R increases with the spectrometer angle (but the resolution get spoiled, as already indicated). Figs.~6 and 7 show that R increases with the incident energy (but again the resolution spoils in that case). 
 
We suggest that the reason for which these narrow, weakly excited structures were not observed till now is due to the lack of experimental precision (resolution and statistics) of previous experiments.

\begin{figure}[!h]
%prl81der.kumac
\begin{center}                                                          
\scalebox{.48}[.48]{
\includegraphics[bb=20 20 530 550,clip=]{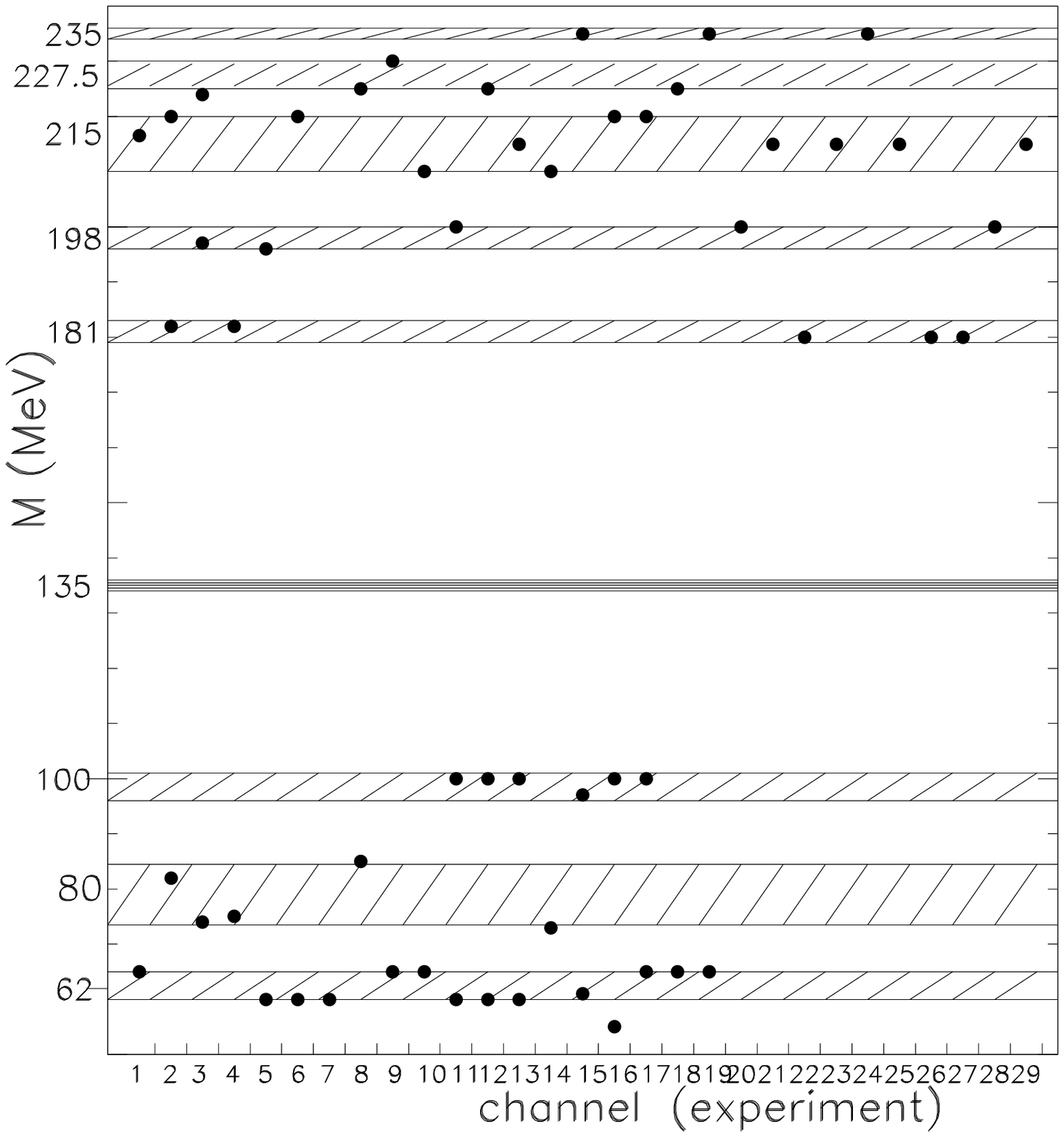}}
\caption{Masses of the weakly excited structures extracted from several experiments.}
\label{fig11}
\end{center}
\end{figure}
\end{document}